\documentclass[twocolumn,showpacs,english,superscriptaddress,preprintnumbers,amsmath,amssymb,floatfix]{revtex4-1}

\usepackage[T1]{fontenc}
\usepackage[latin9]{inputenc}
\usepackage{dcolumn}
\usepackage{bm}
\usepackage{graphicx}
\usepackage{color}
\usepackage{esint}
\usepackage{babel}
\usepackage{amsfonts}
\usepackage{slashed}
\usepackage{enumerate}

\begin{document}

\title{ Kondo physics from quasiparticle poisoning in Majorana devices }

\author{S.~Plugge}
\affiliation{Institut f\"ur Theoretische Physik, Heinrich-Heine-Universit\"at, D-40225  D\"usseldorf, Germany}

\author{A.~Zazunov}
\affiliation{Institut f\"ur Theoretische Physik, Heinrich-Heine-Universit\"at, D-40225  D\"usseldorf, Germany}

\author{E.~Eriksson} 
\affiliation{Institut f\"ur Theoretische Physik, Heinrich-Heine-Universit\"at, D-40225  D\"usseldorf, Germany}
\affiliation{Universit{\'e} Grenoble Alpes, INAC-SPSMS, F-38000, Grenoble, France, and \\ CEA, INAC-SPSMS, F-38000 Grenoble, France}

\author{A.~M.~Tsvelik}
\affiliation{Brookhaven National Laboratory, Upton, NY 11973-5000, USA}

\author{R.~Egger}
\affiliation{Institut f\"ur Theoretische Physik, Heinrich-Heine-Universit\"at, D-40225  D\"usseldorf, Germany}

\date{\today}

\begin{abstract}
We present a theoretical analysis of quasiparticle poisoning in Coulomb-blockaded Majorana fermion systems tunnel-coupled to normal-conducting leads.  Taking into account finite-energy quasiparticles, we derive the effective low-energy theory and present a renormalization group analysis. We find qualitatively new effects when a quasiparticle state with very low energy is localized near a tunnel contact.  For $M=2$ attached leads, such ``dangerous'' quasiparticle poisoning processes cause a spin $S=1/2$ single-channel Kondo effect, which can be detected through a characteristic zero-bias anomaly conductance peak in all Coulomb blockade valleys.  For more than two attached leads, the topological Kondo effect of the unpoisoned system becomes unstable. A strong-coupling bosonization analysis indicates that at low energy the poisoned lead is effectively decoupled and hence, for $M>3$, the topological Kondo fixed point re-emerges, though now it involves only $M-1$ leads. As a consequence, for $M=3$, the low-energy fixed point becomes trivial corresponding to decoupled leads. 
\end{abstract}


\maketitle

\section{Introduction}

Majorana bound states (MBSs) in topological superconductors are presently attracting a lot of attention \cite{Alicea2012,Leijnse2012,Beenakker2013,Franz2015,Beenakker2015}. Recent progress suggests that they can be experimentally realized as end states of topological superconductor (TS) nanowires.  Such TS wires effectively implement the well-known Kitaev chain by contacting helical nanowires (i.e., nanowires with strong spin-orbit coupling in a properly oriented Zeeman field) with bulk $s$-wave superconductors.  Shortly after the first report of MBS signatures via zero-bias anomalies in the tunneling spectroscopy \cite{Mourik2012}, a  second generation of topological nanowires has emerged. These are based on InAs with high-quality proximity coupling to superconducting Al \cite{Chang2015}, which allows one to achieve the hard superconducting proximity gap \cite{Krogstrup2015,Higginbotham2015} needed for the unambigous observation of Majorana fermions.  Evidence for MBSs in such second-generation wires has recently been observed in Coulomb blockade spectroscopy experiments \cite{Albrecht2015}.  Intense experimental efforts are now devoted to elucidating the nonabelian braiding statistics expected for MBSs.  Devices with strong Coulomb effects may be very useful in this regard \cite{Aasen2015}.  
A possible complication in Majorana devices can arise from the presence of low-lying fermionic quasiparticle states.  Many works have studied such ``quasiparticle poisoning''  effects in the absence of topologically protected modes, for instance, see Refs.~\cite{Aumentado2004,Glazman2005,Catelani2011,Brunetti2014}. Given the crucial role of parity conservation for detecting MBS signatures \cite{Alicea2012,Leijnse2012,Beenakker2013}, even a single quasiparticle may drastically affect experimental results for Majorana devices.  Indeed, quasiparticle poisoning has already been analyzed in this context, but only for noninteracting Majorana systems \cite{Rainis2012,Cheng2012,Colbert2014}. 

\begin{figure}[t]
  \centering  
  \includegraphics[width=6cm]{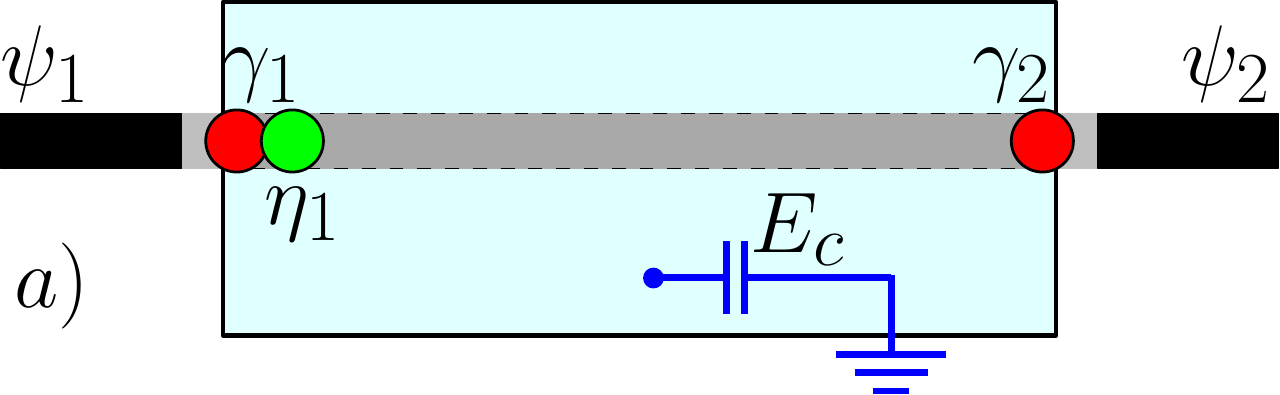}
  
  \bigskip
  
  \includegraphics[width=6cm]{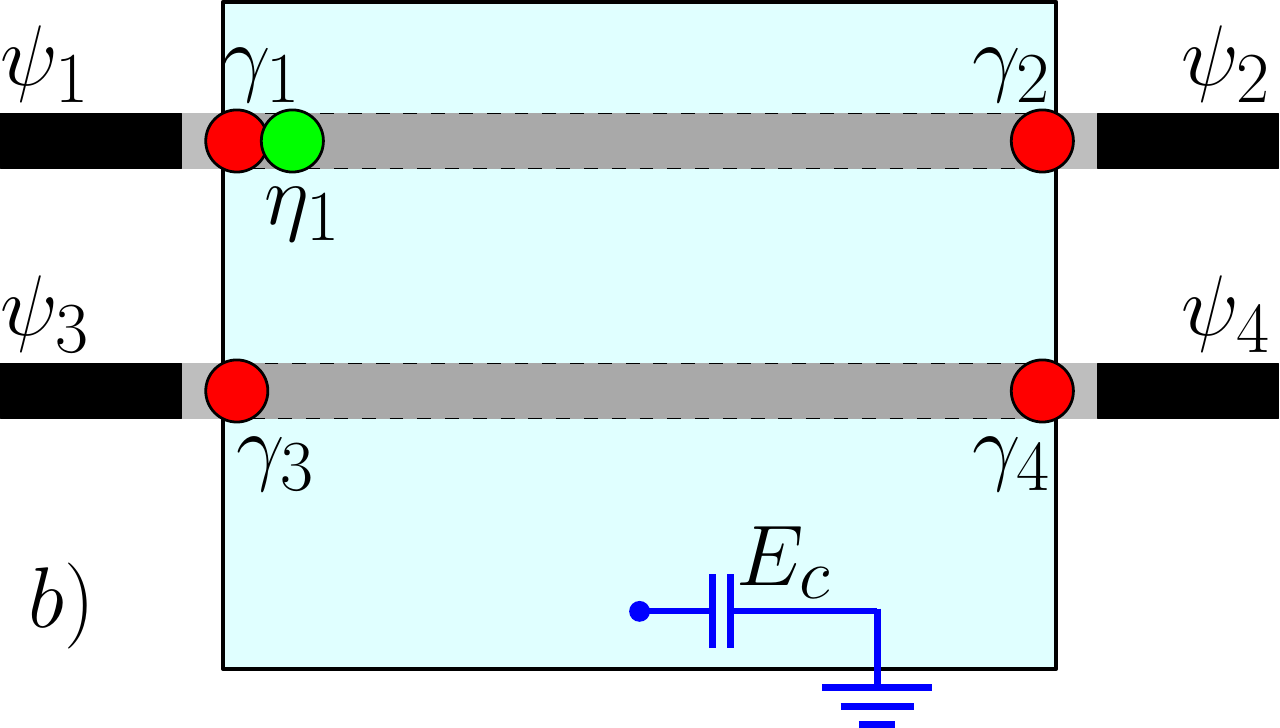}  
  \caption{ (Color online) Schematic setup of the Majorana fermion devices studied here.
A floating mesoscopic superconducting island with large charging energy $E_C$ (light blue box) creates proximity-induced pairing gaps on $N$ adjacent TS wires (grey).  This ``Majorana-Cooper box'' hosts $2N$ MBSs (red circles, corresponding to the Majorana operators $\gamma_j$). To probe transport, $M$ overhanging wire parts (black) serve as normal leads, with electron field operators $\psi_j$.  They are tunnel-coupled to the TS wire ends.  Near one of these ends, a low-energy quasiparticle state is localized (green circle), which effectively corresponds to an extra ``poisoning'' Majorana fermion $\eta_1$, see Sec.~\ref{sec3}.  Panel a) shows the case $N=1$ and $M=2$. Panel b) is for $N=2$ and $M=4$.   
 } \label{fig1}
\end{figure}

In the present work, we instead study the effects of low-lying quasiparticle states in the context of Coulomb-blockaded Majorana devices. The setup is sketched in Fig.~\ref{fig1}.  We consider a floating mesoscopic superconductor with charging energy $E_C$, onto which $N$ helical nanowires have been deposited. Due to the proximity effect, each TS wire hosts a MBS pair.  Below, the island together with the $N$ wires is referred to as ``Majorana-Cooper box'', which  is tunnel-coupled to $M$ normal-conducting leads. The leads could, e.g., be due to non-superconducting ``overhanging'' nanowire parts, see Fig.~\ref{fig1}, where we assume that each TS wire end is contacted by at most one lead, i.e., $M\le 2N$. For nontrivial quantum transport behavior, the minimal case of interest is $M=2$.  Importantly, fermion parity on the box is conserved as long as charge quantization is enforced by a sufficiently large charging energy.

For temperatures well below the proximity gap, one may argue that quasiparticle states are not occupied with significant thermodynamic weight.  Even for a  sub-gap bound state, as long as it is not located near a MBS, the poisoning timescale (on which the occupation of this state will change) should be very long because all matrix elements connecting this quasiparticle state to other low-energy electronic levels, such as MBSs or lead electrons, are small.  A more ``dangerous'' situation arises for sub-gap states located near the TS wire ends, which may occur in practice because the proximity-induced pairing gap $\Delta_w$ also closes there \cite{Tsvelik2012}.  When the TS wires are tunnel-contacted by leads or quantum dots, tunneling processes via the quasiparticle state will then compete with those involving topologically protected MBSs. In order to identify such ``dangerous'' quasiparticle states, it is important to understand this competition and  the resulting physical consequences.  

 Previous work on the setup in Fig.~\ref{fig1} has ignored all quasiparticle states apart from the MBSs. In that case, for $M=2$ attached leads, one arrives at the ``Majorana single-charge transistor'' \cite{Fu2010,Zazunov2011,Hutzen2012}, where the non-locality of the fermion mode built from the two Majorana operators allows for electron teleportation \cite{Fu2010,Sodano2007} and for long-distance entanglement generation between a pair of quantum dots  \cite{Wang2013,Plugge2015}.  For $M>2$ leads, one instead encounters the so-called ``topological Kondo effect'' (TKE)  \cite{BeriCooper2012,AltlandEgger2013,Beri2013,Altland2014,Zazunov2014,Galpin2014,Eriksson2014,Meidan2014,Timm2015,Buccheri2015}.
In spite of charge quantization due to Coulomb blockade, the box ground state is $2^{N-1}$-fold degenerate. This fact can be understood by noting that the Majorana space is a priori $2^{N}$-fold degenerate, but the parity constraint due to charge quantization now removes half of the states.  For $N>1$, the remaining degree of freedom can be viewed as a quantum impurity ``spin'', where the ``real-valuedness'' condition $\gamma^{}_j=\gamma_j^\dagger$ of the Majorana operators implies the symmetry group SO($M)$ instead of SU(2) \cite{BeriCooper2012,Altland2014}.  This spin is effectively exchange-coupled to the lead electrons, and the corresponding screening processes culminate in the TKE, which is of overscreened multi-channel type, represents a non-Fermi liquid fixed point, and is detectable through the temperature dependence of the linear conductance tensor
\cite{BeriCooper2012,AltlandEgger2013,Beri2013,Altland2014,Zazunov2014}.  

Let us now briefly motivate why quasiparticle poisoning is expected to be important for the setup of Fig.~\ref{fig1}. In the absence of poisoning, the in-tunneling of a lead electron into the Majorana-Cooper box has to be followed after a short time $\approx\hbar/E_C$ by the out-tunneling of an electron from the box to some other lead \cite{AltlandEgger2013}. However, with an additional low-energy quasiparticle state present near the TS wire end, the system has a new option: The in-tunneling process can be compensated for by the out-tunneling of a quasiparticle.  Such effects could significantly modify the TKE for $M> 2$, as well as the teleportation or long-range entanglement phenomena for $M=2$. This question is also important in view of the fact that the Majorana-Cooper box is a basic building block in Majorana surface code proposals \cite{Terhal2012,Vijay2015,Landau2015}. In these proposals, the effective quantum impurity spins (each of which is encoded by one Majorana-Cooper box) are arranged on  a two-dimensional lattice, which then is employed for quantum information processing.   

The structure of the remainder of this paper is as follows.  In Sec.~\ref{sec2}, we model  the setup in Fig.~\ref{fig1}, with an emphasis on new aspects introduced by  quasiparticle poisoning. In the absence of poisoning, our Hamiltonian below reduces to previously studied models.  For clarity, we mainly focus on the case of a single relevant  quasiparticle state of energy $E_\Delta\ll E_C$. We derive the effective low-energy theory, $H_{\rm eff}$, by a Schrieffer-Wolff transformation in Sec.~\ref{sec3a}.   For $E_\Delta\gg k_B T_K$, we predict an enhancement of the Kondo temperature $T_K$ for the TKE, see Sec.~\ref{sec3b}. Quasiparticle poisoning thus is not necessarily detrimental to the observation of this non-Fermi liquid state: it may actually help to access the $T\ll T_K$ regime.   In Sec.~\ref{sec4}, we turn to the simplest $M=2$ case, where despite of the effectively spinless nature of the system,  $H_{\rm eff}$ is equivalent to the anisotropic (XYZ) spin $S=1/2$ single-channel Kondo model \cite{Gogolin1998}, which flows to an isotropic Fermi liquid strong-coupling fixed point on energy scales below $k_B T_K^{(M=2)}$. We determine the respective Kondo temperature, $T_K^{(M=2)}$, and discuss the zero-bias anomaly conductance peak caused by the many-body Kondo resonance.   Next, in Sec.~\ref{sec5}, we turn to the case of arbitrary $M$. In Sec.~\ref{sec5a}, we apply  Abelian bosonization  \cite{Gogolin1998} to study the most challenging case $E_\Delta\ll k_B T$, see also App.~\ref{appa}.  In Sec.~\ref{sec5b}, we determine the perturbative renormalization group (RG) equations, cf.~App.~\ref{appb}, and we show that the TKE is destabilized by dangerous quasiparticle poisoning processes. However, the strong-coupling analysis presented in Sec.~\ref{sec6} shows that for $M>3$, a TKE with symmetry group SO($M-1$) re-emerges at low temperatures. The effective change $M\to M-1$ is rationalized by noting that only $M-1$ leads (those not attached to the TS wire end that hosts the poisoning quasiparticle) will contribute to the low-energy sector.  For $M=3$, the RG flow instead proceeds to a fixed point corresponding to effectively decoupled leads. We finally present our conclusions in Sec.~\ref{sec7}.  Throughout the paper, we employ units where $\hbar=1$. 

\section{Model}\label{sec2}

In this paper, we present a theoretical analysis for the low-energy transport properties of the generic setup in Fig.~\ref{fig1}. The central element of the setup is the Majorana-Cooper box, where $N$ nanowires are in proximity to the same floating mesoscopic superconductor. When driven into the topologically nontrivial phase,  each of these TS wires hosts a pair of zero-energy Majorana end states \cite{Alicea2012,Alicea2011}.  We shall assume sufficiently long TS wires such that the hybridization between different MBSs can be neglected; for a discussion of these effects, see, e.g., Ref.~\cite{Altland2014}. 
Recent experiments have shown that this requirement can be fulfilled for available InAs/Al nanowires \cite{Albrecht2015}.   The box is then connected to $M$ (with $M\le 2N$) normal-conducting leads by tunnel couplings, see Fig.~\ref{fig1}.  The Hamiltonian is thereby written as 
\begin{equation}\label{totham}
H=H_c+H_{qp}+H_{leads}+H_t,
\end{equation}
where $H_c$ captures Coulomb charging effects, $H_{qp}$ models finite-energy quasiparticles in the TS, $H_{leads}$ describes the $M$ normal-conducting leads, and $H_t$ is a tunneling Hamiltonian connecting the box to the leads.  We next describe these contributions.

In concrete realizations, the leads may be defined by the ``overhanging'' non-superconducting wire parts, see Fig.~\ref{fig1}. We model them as semi-infinite one-dimensional (1D) channels of noninteracting spinless (helical) fermions.  For the case of point-like tunneling studied below, this model also describes transport for bulk (2D or 3D) electrodes \cite{Zazunov2011,Fidkowski2012}.  With the coordinate $x\le 0$ for a given lead ($j=1,\ldots,M$), we have a pair of right- and left-movers in each wire, $\psi_{j,R/L}(x)$, and the electron field operator is $\psi_j(x)=e^{ik_Fx}\psi_{j,R}(x)+ e^{-ik_Fx}\psi_{j,L}(x)$, where $k_F$ is the Fermi momentum. On low energy scales, the generic lead Hamiltonian now takes the form
\begin{equation}\label{Hleads}
H_{leads} = -i v_F \sum_{j=1}^M  \int_{-\infty}^0 dx \left( \psi^\dagger_{j,R} \partial_x\psi^{}_{j,R} -\psi^\dagger_{j,L} \partial_x\psi^{}_{j,L}
\right),
\end{equation}
where $v_F$ denotes the Fermi velocity. At $x=0$, the boundary conditions $\psi_{j,R}(0)=\psi_{j,L}(0)$ are enforced, and point-like tunneling processes involve the lead operators $\psi^{}_j(0)$ and $\psi_j^\dagger(0)$.   

Turning to the Hamiltonian of the Majorana-Cooper box, $H_{c}+H_{qp}$, we first note that the proximity-induced pairing gap in the TS wires is typically well below the bulk superconducting gap  \cite{Alicea2012}. We therefore take into account quasiparticles only in the TS wires, which are either continuum (above-gap) states (cf.~also Sec.~\ref{sec7}) or localized (sub-gap) bound states.  Each TS wire corresponds to a $p$-wave superconductor, where the proximity-induced gap profile $\Delta_w(x)$ can be chosen real-valued in a suitable gauge \cite{Zazunov2012}. With Fermi velocity $v_0$, the single-particle Bogoliubov-de Gennes (BdG) Hamiltonian for a given wire reads \cite{Alicea2012,Leijnse2012}
\begin{equation}\label{BdGH}
{\cal H}_{BdG} = - i v_0 \sigma_z \partial_x   + \sigma_x \Delta_w(x),
\end{equation}
where the Pauli matrices $\sigma_{a=x,y,z}$ act in particle-hole (Nambu) space. Particle-hole and conjugation symmetry properties are expressed by  
\begin{equation}\label{symrel}
\sigma_z {\cal H}_{BdG}^\ast \sigma_z = \sigma_y {\cal H}_{BdG} \sigma_y = - {\cal H}_{BdG}.
\end{equation}  
For eigenstates of the BdG equation, ${\cal H}_{BdG}\Phi_E= E  \Phi_E$,  Eq.~\eqref{symrel} implies the symmetry relations
\begin{equation}\label{uv}
\Phi_E(x) = \left( \begin{array}{c} u_E(x) \\ v_E(x) \end{array} \right) = \sigma_z \Phi_{-E}^\ast(x) ,\quad u^\ast_E(x) = v_E(x) .
\end{equation}
The last relation in Eq.~\eqref{uv} can be rationalized by noticing that in a rotated  basis, $\sigma_z \rightarrow e^{i \sigma_x \pi/4} \sigma_z  e^{-i \sigma_x \pi/4} = \sigma_y$, the BdG equation admits purely real solutions.  

We now switch to a second-quantized formulation and introduce the Nambu field operator for a given TS wire, 
$\Psi(x)=\left(\psi^{}_R(x), \psi_{L}^\dagger(x)\right)^T,$
where $\psi_{R/L}$ refer to left/right-moving field operators in the TS. With the BdG single-particle Hamiltonian (\ref{BdGH}),  the quasiparticle Hamiltonian for a single wire follows in the form
\begin{equation}\label{Hqp0}
H_{qp}^{(1)} =\int dx \, \Psi^\dagger(x) {\cal H}_{BdG} \Psi(x).
\end{equation} 
For $E>0$, let us now define conventional fermion operators $f_{e,E}$  $\left( f_{h,E} \right)$ for particle-like (hole-like) excitations of energy $E$ $(-E)$. Taking into account Eq.~\eqref{uv}, $\Psi(x)$ then has the mode expansion 
\begin{eqnarray}
\Psi(x) =\Psi_{\rm MBS}(x) &+&  \sum_{E>0} \left| u_E(x) \right| \Biggl[ e^{i \sigma_z \zeta_E(x)/2} \left( \begin{array}{c} 1 \\ 1 \end{array} \right) f_{e,E}
 \nonumber \\ &+&   \ e^{-i \sigma_z \zeta_E(x)/2} \left( \begin{array}{c} 1 \\ -1 \end{array} \right) f_{h,E}^\dagger \Biggr] ,\label{modeexp}
\end{eqnarray}
where the real-valued phase $\zeta_E(x)$ follows by solving the BdG equation and the energy summation extends over positive BdG eigenvalues. 
The MBS contribution $\Psi_{\rm MBS}$ will be taken into account in the tunneling Hamiltonian below.  However, zero-energy modes do not contribute to
the quasiparticle Hamiltonian (\ref{Hqp0}). Inserting Eq.~\eqref{modeexp} into Eq.~\eqref{Hqp0}, and subsequently summing over all TS wires ($\alpha=1,\ldots,N$), we obtain 
\begin{equation}\label{Hqp}
H_{qp} =  \sum_{\alpha,E>0} E \left( f_{\alpha, e, E}^\dagger f^{}_{\alpha,e,E}+ f_{\alpha, h, E}^\dagger f^{}_{\alpha,h,E}\right).
\end{equation}
In addition to the finite-energy quasiparticles just discussed, we also have $2N$ zero-energy MBSs on the box, which are localized near the TS wire ends. 
 They correspond to a set of self-adjoint operators, $\gamma_j = \gamma_j^\dagger$, subject to the anticommutator algebra  
$\left\{ \gamma_j, \gamma_k \right\} = \delta_{jk}$.  A pair of Majorana operators defines a fermion \cite{Alicea2012,Leijnse2012,Beenakker2013}, e.g., 
$d_\alpha = \left( \gamma_{2\alpha-1} + i \gamma_{2\alpha} \right) / \sqrt{2}$ for each TS wire, where the Majorana algebra implies standard fermion anticommutation rules for the $d_\alpha^{}$ and $d_\alpha^\dagger$ operators.  

Turning to the interaction contribution, with the single-electron charging energy $E_C$ and a dimensionless backgate parameter $n_g$, capacitive Coulomb charging effects on the box are contained in   \cite{Fu2010,Zazunov2011,Zazunov2012}
\begin{equation}\label{Hc}
H_c= E_C \left( 2\hat N_c + \hat n-n_g \right)^2,
\end{equation}
where $\hat N_c$ is the Cooper pair number operator and $\hat n$ counts both the occupation of states in the zero-energy Majorana sector and of finite-energy quasiparticle states,
\begin{equation}
\hat n = \sum_\alpha  d^\dagger_\alpha d^{}_\alpha + \sum_{\alpha, E>0} \left( f_{\alpha, e, E}^\dagger
 f^{}_{\alpha,e,E}+ f_{\alpha, h, E}^\dagger f^{}_{\alpha,h,E}\right).
\end{equation} 
The respective eigenvalues $N_c$ and $n$ take integer values.  We note that the condensate phase $\chi$ is conjugate to 
$2 \hat N_c$, with the canonical commutator $\left[ \chi, 2 \hat N_c \right] = i$. As a consequence, the operator $e^{-2 i \chi}$ annihilates one Cooper pair.

From now on, we shall assume that only a single quasiparticle state with energy $E=E_\Delta\ll E_C$ is relevant, plus the hole state required by particle-hole symmetry. Apart from its simplicity, this minimal case is also of considerable practical interest: Noting that the gap function $\Delta_w(x)$ vanishes near the TS wire ends, an 
important example for such a low-energy quasiparticle comes from sub-gap bound states that may be formed near a tunnel contact \cite{Tsvelik2012}. 
While the model and the techniques used here can be directly extended to 
the case of many low-energy quasiparticles, the effects of different quasiparticles are not simply additive in this interacting system, cf.~the discussion before Eq.~\eqref{fourpar} in Sec.~\ref{sec5b}.
Taking into account only one quasiparticle state, say, near the left end of TS wire $\alpha=1$, cf.~Fig.~\ref{fig1}, Eq.~\eqref{Hqp} simplifies to
\begin{equation}\label{Hqp1}
H_{qp} = E_\Delta \left( f_e^\dagger f_e^{} + f_h^\dagger f_h^{} \right) .
\end{equation}
The quasiparticle state also couples to electrons in normal lead $j=1$ by tunneling processes encoded in the tunneling Hamiltonian $H_{t,qp}$. We here assume pointlike tunneling, $H_{t} \sim \psi_1^\dagger(0) [\psi_L(0) + \psi_R(0)] + {\rm h.c.}$, and 
for the moment ignore charge conservation issues. Employing Eq.~\eqref{modeexp} and the definition of the Nambu spinor, we obtain
\begin{equation}\label{Htqp}
H_{t,qp}  =t_1 \psi_1^\dagger(0) \left( e^{i \zeta/2} \eta_e - i e^{-i \zeta/2} \eta_h \right) + {\rm h.c.} ,
\end{equation}
with the generally complex-valued tunnel amplitude $t_1$.  The Majorana operators $\eta_{e/h}^{} = \eta_{e/h}^\dagger$ are built from quasiparticle fermion operators,
\begin{equation}\label{etadef1}
\eta_e = \left( f_e^{} + f_e^\dagger \right)/\sqrt{2} ,\quad\eta_h = -i \left( f_h^{} - f_h^\dagger \right)/\sqrt{2} .
\end{equation}
As a consequence of particle-hole symmetry, see Eqs.~\eqref{symrel} and \eqref{uv}, and the assumption of pointlike tunneling, the lead fermion $\psi_1$ therefore couples to the 
``hybrid'' fermion $\eta_e + i \eta_h$ (and its conjugate).  Taking into account also the tunneling processes involving the 
topologically protected Majorana fermions $\gamma_j$, cf.~Refs.~\cite{Fu2010,Zazunov2011}, we arrive at the tunneling Hamiltonian  
\begin{equation}\label{Ht}
H_t = \sum_{j=1}^M \lambda_j \left( \psi_j^\dagger(0) \gamma_j  + {\rm h.c.} \right)  + H_{t,qp}.
\end{equation}
Without loss of generality, the tunnel amplitudes $\lambda_j$ connecting MBSs to the respective leads can be taken real-valued and positive \cite{Zazunov2012}. 

At this stage, we pause to incorporate the charge conservation condition for our floating (not grounded) device. 
As has been shown in Ref.~\cite{Zazunov2011}, this condition can be taken into account by the following steps. First, for $H_{t,qp}$ in Eq.~\eqref{Htqp}, we put 
\begin{equation}
\psi_1^\dagger \left (f^{}_{e/h} \pm f_{e/h}^\dagger \right)\to 
\psi_1^\dagger \left( f^{}_{e/h} \pm e^{-2 i \chi} f_{e/h}^\dagger \right),
\end{equation}
such that ``anomalous'' processes $\sim \psi_1^\dagger f_{}^\dagger$ will be accompanied by the splitting of a Cooper pair, which in turn is 
implemented by the operator $e^{-2i\chi}$. Second, after rewriting the Majorana operators $\gamma_j$ in terms of $d_\alpha^{}$ and $d_\alpha^\dagger$ fermions, a similar replacement is performed in the Majorana part of Eq.~\eqref{Ht}.  With these changes, $H$ is explicitly charge conserving.   

Finally, in order to arrive at maximally transparent expressions, we remove the $\hat n$ term in the charging 
contribution $H_c$, see Eq.~\eqref{Hc}, by a gauge transformation, 
$H \to e^{i \chi \hat n} H e^{-i \chi \hat n}.$
Using $e^{i \chi \hat n} f_{e/h} e^{-i \chi \hat n} = e^{-i \chi} f_{e/h}$, and similarly for the $d_\alpha$ fermions, we find that  $H_{leads}$ is still given by Eq.~\eqref{Hleads} and $H_{qp}$ by Eq.~\eqref{Hqp1}. The charging energy term reads
\begin{equation}\label{Hc2}
H_c = E_C \left( \hat Q - n_g \right)^2  ,
\end{equation}
where the charge operator $\hat Q$ has integer eigenvalues $Q$, with canonical  commutator $[\chi,\hat Q]=i$. This implies that the operator $e^{i\chi}$ ($e^{-i\chi}$) adds (removes) charge $e$ to (from) the box. 
The tunneling Hamiltonian now takes the form 
\begin{equation}\label{Ht1}
H_t  =  \sum_{j=1}^M \lambda_j \left( \psi_j^\dagger(0) e^{-i \chi} \gamma_j + {\rm h.c.}\right) + H_{t,qp}, 
\end{equation}
with the quasiparticle tunneling contribution
\begin{equation}\label{Ht11}
H_{t,qp} =  t_1 \psi_1^\dagger(0) e^{-i \chi}
\left( e^{i \zeta/2} \eta_e - i e^{-i \zeta/2} \eta_h \right)  + {\rm h.c.}
\end{equation}

\section{Effective low-energy Hamiltonian}\label{sec3}

\subsection{Schrieffer-Wolff transformation} \label{sec3a}

In this section, we derive an effective low-energy description for the general model discussed in Sec.~\ref{sec2}, which holds under the following conditions. First, we take into account only one quasiparticle state at energy $E_\Delta$, localized near a TS wire end. (We briefly discuss the case of delocalized above-gap quasiparticles in Sec.~\ref{sec7}.) Second, the charging energy should be the dominant energy scale,
\begin{equation} \label{cond1}
E_C\gg {\rm max}(k_B T, E_\Delta, \lambda^2_j/v_F, |t_1|^2/v_F).
\end{equation} 
Third, we assume that $n_g$ is close to an integer.  In the regime defined by Eq.~\eqref{cond1}, the system then exhibits charge quantization, $Q=[n_g]$. 

According to Eq.~\eqref{Ht11}, the lead fermion $\psi_1$ is tunnel-coupled to three Majorana fermions, namely to the topologically protected Majorana operator $\gamma_1$ and 
to the Majorana fermions $\eta_{e/h}$ describing the real and imaginary parts of the quasiparticle operators $f_{e}$ and $f_h$, resp., see Eq.~\eqref{etadef1}. Through the 
quasiparticle Hamiltonian $H_{qp}$ in Eq.~\eqref{Hqp1}, $\eta_{e}$ and $\eta_h$ also couple with strength $E_\Delta$ to the two additional Majorana fermions representing the imaginary and real part of $f_{e}$ and $f_h$, respectively.  However, even though this $E_\Delta$-coupling constitutes a relevant perturbation in the RG sense, it does not affect the scaling properties of the system for temperatures within the window
\begin{equation} \label{cond2}
E_\Delta\ll k_B T\ll E_C.
\end{equation}  
For the sake of clarity, we will mainly focus on the regime defined by Eqs.~\eqref{cond1} and \eqref{cond2}, where one can effectively put $E_\Delta\to 0$, with $\zeta\to 0$ in Eq.~\eqref{Ht11}. (The case $E_\Delta\gg k_B T$ will be separately addressed in Sec.~\ref{sec3b}.)

Next, it is beneficial to switch to new Majorana operators $\tilde\gamma_1$ and $\eta_1$, representing linear combinations of $\gamma_1$, $\eta_{e}$ and $\eta_{h}$.   
This step allows us to decouple one of these three Majorana fermions from the problem.  To that end, using Eq.~\eqref{Ht1} and writing $t_1=|t_1|e^{i\varphi}$, we define
\begin{eqnarray}\nonumber
\tilde\lambda_1 \tilde\gamma_1 & = & \lambda_1 \gamma_1 + |t_1| \left [\cos(\varphi) \eta_e+ \sin(\varphi) \eta_h \right], \\ \label{newmbs} 
\tilde t \eta_1 &=&  |t_1| \left[-\sin(\varphi) \eta_e+\cos(\varphi) \eta_h \right],
\end{eqnarray}
where $\tilde\lambda_1=\sqrt{\lambda_1^2+|t_1|^2}$ and $\tilde t=|t_1|$ are new (real-valued positive) tunnel couplings.  Equation \eqref{newmbs} is evidently consistent with the Majorana operator algebra, in particular $\{ \tilde\gamma_1, \eta_1\}=0$.
In order to simplify the notation, we finally rename $\tilde \gamma_1\to \gamma_1$, as well as the respective MBS coupling to the lead, $\tilde\lambda_1\to \lambda_1$. 
The tunneling Hamiltonian for the contact to lead $j=1$ is therefore given by
\begin{equation}\label{Ht5}
H_t^{(j=1)} = \psi_1^\dagger(0) e^{-i \chi} \left( \lambda_1 \gamma_1 - i \tilde t \eta_1 \right) + {\rm h.c.}, 
\end{equation}
where it is worth stressing that, in effect, only a single ``poisoning'' Majorana fermion ($\eta_1$) remains in the problem.

We now employ a standard Schrieffer-Wolff transformation \cite{BravyiLoss2011} to project the system to the low-energy Hilbert space
spanned by states with quantized box charge $Q=[n_g]$.  This projection takes into account virtual excitations of higher-order 
charge states and has been described for the same system in the absence of poisoning in Ref.~\cite{BeriCooper2012}.  
Including poisoning effects, we now arrive at the effective low-energy Hamiltonian $H_{\rm eff} = H_{leads}+H_K$, where 
\begin{eqnarray}\label{Hint}
H_{K} &=& \sum_{(j \neq k)=1}^M J_{jk} \psi_j^\dagger(0) \psi^{}_k(0) \gamma_k \gamma_j \\ \nonumber &+&
i \sum_{j=1}^M K_{j1} \left( \psi_j^\dagger(0) \psi^{}_1(0) \eta_1 \gamma_j - {\rm h.c.} \right) ,
\end{eqnarray}
with real-valued non-negative ``exchange couplings'' 
\begin{equation}\label{exchange}
J_{jk} = \frac{2 \lambda_j \lambda_k }{ E_C}, \quad K_{jk} = \frac{2 \lambda_j \tilde t}{ E_C} \delta_{k,1}.
\end{equation}
For $M\ge 3$, the first term in Eq.~\eqref{Hint} reduces to the TKE model \cite{BeriCooper2012}. Indeed, in the absence of $\eta_1$, the low-energy box degrees of freedom correspond to a ``spin'' operator of symmetry group SO$(M)$, which has the components $i\gamma_j\gamma_k$ \cite{Altland2014} and is exchange-coupled to a lead electron ``spin'' density at $x=0$, see Eq.~\eqref{Hint}. The exchange couplings $J_{jk}$ are marginally relevant under RG transformations and scale towards an isotropic strong-coupling fixed point describing the TKE.  The second term in Eq.~\eqref{Hint} is new and describes additional exchange-type couplings involving the poisoning Majorana fermion $\eta_1$.  

Together with $H_{leads}$ in Eq.~\eqref{Hleads}, the Hamiltonian (\ref{Hint}) defines our low-energy model for quasiparticle poisoning in a Majorana device operating under strong Coulomb blockade conditions. In this model, we consider a quasiparticle state localized near one tunnel contact, such that effectively several MBSs will be tunnel-coupled to the same lead.  Similar but different models have also been studied recently by others \cite{Meidan2014,Timm2015}. As discussed below, this modification of the clean TKE has interesting consequences that may be observable in Coulomb spectroscopy experiments. 

\subsection{Intermediate quasiparticle energy}\label{sec3b}

Before studying $H_{\rm eff}=H_{leads}+H_K$ through a bosonization analysis, let us briefly turn to the regime of intermediate quasiparticle energy, $k_B T \ll E_{\Delta} \ll E_C$, where an effective low-energy theory only involving the topologically protected Majorana fermions $\gamma_j$ is applicable.  Indeed, in this regime, since occupation of the $f_{e,h}$ quasiparticle states now comes with the large energy cost $E_\Delta$, see $H_{qp}$ in Eq.~\eqref{Hqp1}, we can project $H_{\rm eff}$ also to the ground-state sector of $H_{qp}$. 

  Using the tunneling Hamiltonian $H_t$ in Eqs.~\eqref{Ht1} and \eqref{Ht11}, we first perform a Schrieffer-Wolff transformation in order to project away the higher-order charge states. Subsequently, since we are interested in energy scales well below $E_\Delta$, we also project to the ground-state sector of $H_{qp}$ by a second Schrieffer-Wolff transformation.  
The resulting low-energy Hamiltonian is given by $H_{\rm eff}=H_{leads}+\tilde H_K$, with
\begin{equation} \label{heffe3}
\tilde H_{K} =  \sum_{j \ne k} \tilde J_{jk} \psi_j^{\dagger}\psi^{}_k   \gamma_k \gamma_j ,
\end{equation}
plus an RG-irrelevant potential scattering term $\sim \psi_1^{\dagger} \psi_1^{}$.  (We note that $\gamma_1$ here refers to the ``original'' Majorana operator, without the transformation in Eq.~\eqref{newmbs}.) When compared to the small-$E_\Delta$ exchange term [$H_K$ in Eq.~\eqref{Hint}], 
we observe that all terms related to the quasiparticle state have disappeared, except for a renormalization of the couplings $J_{jk}$. Instead of 
Eq.~\eqref{exchange}, which gives $J_{jk}$ already in the absence of poisoning, we now find $J_{jk}\to \tilde J_{jk}$ with
\begin{equation}\label{increaseJ}
\tilde J_{jk} =  J_{jk} \left(1 +   \frac{2  |t_1|^2 }{ E_{\Delta} E_C }  \right).
\end{equation}
We mention in passing that for several quasiparticles with energy above $k_B T$, Eq.~\eqref{increaseJ} simply acquires independent corrections of the form quoted here.  
Tunneling processes via the quasiparticle state  ($ t_1\ne 0$) therefore increase the $J$ couplings, which can be rationalized by noting that an additional channel for cotunneling processes through the box has now become available.  This channel is due to the high-energy quasiparticle state. The increase $J\to \tilde J$ then implies an upward renormalization of the Kondo temperature $T_K$ characterizing the TKE for $M \ge 3$. For isotropic couplings, one finds $k_B T_K \approx E_C  e^{-1/[\nu_0(M-2) J]}$, with the lead density of states $\nu_0$ \cite{BeriCooper2012}.  
We conclude that quasiparticle poisoning will not necessarily destroy the TKE.  To the contrary, when a quasiparticle state is localized near a tunnel contact and has energy $E_\Delta> k_B T_K$,  access to the $T\ll T_K$ regime becomes easier through the described $T_K$ enhancement mechanism.

\section{Conventional Kondo physics:  $\bm{M=2}$} \label{sec4}

From now on, we shall discuss the more challenging case of a low-energy quasiparticle, where we can effectively put $E_\Delta\to 0$.  The simplest scenario considers $M=2$ attached leads, which we discuss in this section. For $M=2$, there are only three independent exchange couplings, 
\begin{equation}\label{initRG}
J_x= 2K_{21}, \quad J_y= 2J_{21},\quad J_z= -2K_{11}.
\end{equation}
Their bare (initial) values follow from Eq.~\eqref{exchange}.  The effective low-energy Hamiltonian, $H_{\rm eff}=H_{leads}+H_K$ with $H_K$ in Eq.~\eqref{Hint}, is then equivalent to the fully anisotropic $S=1/2$ single-channel Kondo model. To establish this correspondence, we introduce a $S=1/2$ ``quantum impurity spin'' operator with components
\begin{equation}
S_x = i\eta_1\gamma_2, \quad S_y =i\gamma_2\gamma_1, \quad S_z =i \gamma_1\eta_1.
\end{equation}
Noting that the coupling of this spin operator to the identity operator, $\sum_{j=1,2}\psi_j^\dagger(0)\psi_j^{}(0)$, does not generate RG-relevant scaling operators, and taking into account the exchange couplings defined in Eq.~\eqref{initRG}, we find that $H_K$ in Eq.~\eqref{Hint} is equivalent to the fully anisotropic (XYZ) exchange term 
\begin{eqnarray} \label{XYZ}
H_{K} &=& J_x S_x s_x + J_y S_y s_y + J_z S_z s_z,\\ 
\nonumber s_a &=& \frac12 \sum_{j,k=1,2}  \psi^\dagger_j (0) \sigma^a_{jk}\psi^{}_k(0),
\end{eqnarray}
with Pauli matrices $\sigma^{a}$ in lead space. 

The anisotropic $S=1/2$ single-channel Kondo model can be solved by the Bethe ansatz \cite{caveat}. The model scales towards a strong-coupling Fermi liquid fixed point \cite{Gogolin1998}, where the exchange couplings $J_a$ become more and more isotropic. In order to obtain the Kondo temperature, $T_K^{(M=2)}$, determining the crossover scale from weak to strong coupling, we consider the standard RG equations for this problem. With the flow parameter $\ell$, with $d\ell=d\ln\tau_c$ for running short-time cutoff $\tau_c$ \cite{Cardy1996}, and the couplings in Eq.~(\ref{initRG}), we arrive at the symmetric RG equations
 \begin{equation}\label{M2rge}
\frac{dJ_x}{d\ell}  = J_y J_z,\quad \frac{dJ_y}{d\ell}  = J_z J_x,\quad \frac{dJ_z}{d\ell}  = J_x J_y.
\end{equation}  
It is straightforward to show from Eq.~\eqref{M2rge} that two invariants during the RG flow are given by 
\begin{equation} \label{invariants}
{\cal I}_1=J_x^2-J_y^2, \quad {\cal I}_2=J_x^2-J_z^2.
\end{equation}
Under the assumption $J_x^2 > J_z^2 > J_y^2$, Eq.~\eqref{M2rge} yields 
\begin{equation}\label{reducedeq} 
\frac{dJ_x}{d\ell} = \sqrt{(J_x^2 - {\cal I}_1)(J_x^2 - {\cal I}_2)}.
\end{equation}
By integration of Eq.~\eqref{reducedeq}, we then extract the Kondo temperature as the scale at which $J_x(\ell)$ diverges,  
\begin{equation}\label{tkm2}
k_BT^{(M=2)}_K = D\exp\left( - \frac{F\left(\sin^{-1}\left[\frac{\sqrt{{\cal I}_1}}{J_x(0)}\right],\sqrt{\frac{ {\cal I}_2}{{\cal I}_1} } \ \right) }{\nu_0\sqrt{{\cal I}_1}}\right),
\end{equation}
where $F(\phi,k)$ is the elliptic integral of the first kind \cite{Gradsteyn},  $D$ denotes the bandwidth, and $\nu_0$ is the lead density of states.  For almost isotropic initial conditions, Eq.~\eqref{tkm2} can be simplified and reduces to the more familiar expression $k_B T_K^{(M=2)}\simeq D e^{-1/[\nu_0 J_x(0)] }$.

For $T\ll T_K^{(M=2)}$, the spin $S=1/2$ single-channel Kondo fixed point will be approached, where deviations from isotropy are dynamically suppressed.   The low-temperature behavior thus corresponds to conventional $S=1/2$ Kondo physics, where the formation of a many-body Kondo resonance allows for resonant tunneling through the Majorana-Cooper box.  The predicted Kondo physics should be experimentally observable in setups similar to the one of Ref.~\cite{Albrecht2015} through a narrow conductance peak of width $T_K^{(M=2)}$ around zero bias voltage (``Kondo ridge''). The linear conductance between leads 1 and 2 then approaches the quantized value $e^2/h$ for $T\ll T_K^{(M=2)}$, see Ref.~\cite{Gogolin1998},
\begin{equation}\label{kondocondu}
G_{12}(T) = \frac{e^2}{h}\left[ 1- c_2 \left(T/T_K^{(M=2)}\right)^2 \right],
\end{equation}
with a coefficient $c_2$ of order unity. The temperature dependence of the conductance here follows from Fermi liquid theory.  Importantly, these Kondo ridges are predicted to appear in \textit{all}\ Coulomb valleys, in contrast to conventional quantum dots where they are found in ``odd'' valleys only \cite{Gogolin1998}. 

We conclude that for $M=2$, ``dangerous'' quasiparticle poisoning processes are responsible for a $S=1/2$ single-channel Kondo effect.
The resulting conductance peak structure can easily be distinguished from standard Kondo features due to the electronic spin in quantum dots,
as well as from the resonant Andreev reflection peaks found in noninteracting (grounded) Majorana devices \cite{Alicea2012,Leijnse2012,Beenakker2013} which are independent of the backgate parameter $n_g$. 

\section{Topological Kondo effect and quasiparticle poisoning }\label{sec5}

\subsection{Abelian bosonization}\label{sec5a}

In order to discuss the general case of $M>2$ leads,  it is convenient to employ Abelian bosonization for the lead fermions \cite{Fidkowski2012,Gogolin1998}.   Within this approach, the 1D fermion operators have the equivalent bosonized form
\begin{equation}\label{fieldop}
\psi_{j,R/L}(x)=a_c^{-1/2} \Gamma_j e^{i[\phi_j(x)\pm \theta_j(x)]}, 
\end{equation}
where $a_c$ is a microscopic short-distance lengthscale. The dual pairs of boson fields ($\phi_j,\theta_j)$ have the commutator algebra 
$\left[ \phi_j(x) , \theta_k (x') \right] = i (\pi/2) {\rm sgn}(x-x') \delta_{jk}$.
Equation \eqref{fieldop} also makes use of auxiliary Majorana operators $\Gamma_j^{}=\Gamma_j^\dagger$ with the anticommutator algebra $\left\{ \Gamma_j, \Gamma_k \right\} = \delta_{jk}$, which represent the Klein factors needed to ensure anticommutation relations for fermions on different wires. This Klein  factor representation allows for significant technical advantages in Majorana devices \cite{AltlandEgger2013,Beri2013}. The lead Hamiltonian (\ref{Hleads}) has the bosonized form
 \begin{equation}\label{H0}
H_{leads} =  \sum_{j=1}^M \frac{v_F}{2 \pi} \int_{-\infty}^0 dx \left[\left( \partial_x \phi_j \right)^2 + \left( \partial_x \theta_j \right)^2\right],
\end{equation}
where $\psi_{j,R}(0)=\psi_{j,L}(0)$ yields the boundary conditions $\theta_j(0)=0$.

 In order to bosonize $H_K$ in Eq.~\eqref{Hint}, it is convenient to employ the shorthand notation
\begin{equation}\label{shorthand}
\Phi_j=\phi_j(0),\quad \Theta'_j=\partial_x\theta_j(0), \quad {\rm i.e.,} \quad \psi_j(0)\sim \Gamma_j e^{i\Phi_j}.
\end{equation} 
As we show in App.~\ref{appa}, by combining the physical Majorana fermions ($\gamma_j$ and $\eta_1$) with the Majorana fermions $\Gamma_j$ representing the Klein factors,  parity conservation allows one to efficiently capture the dynamics of all these Majorana fermions, for arbitrary number of leads $M$, by just a single ``pseudospin'' $S=1/2$ operator with components $\tau_{a}/2$ (where $a=x,y,z$). Following the steps in App.~\ref{appa},  the bosonized form of $H_K$ is given by
\begin{eqnarray}\label{Hint4}
H_K &=& \sum_{1<j < k}^M J_{jk} \cos\left(\Phi_j - \Phi_k\right) +\sum_{j=1}^M K_{jj} \tau_z \Theta'_j  \\ \nonumber &+&
\sum_{j = 2}^M \left[  \left( L_{j1}^{(+)} \tau_+ + L_{j1}^{(-)}\tau_-\right)e^{i(\Phi_j - \Phi_1)} + {\rm h.c.}
\right],
\end{eqnarray}
where $\tau_\pm = (\tau_x \pm i \tau_y)/2$. We note that a factor $1/a_c$ has been absorbed in the exchange couplings $J_{jk}$ and $K_{jk}$, see Eq.~\eqref{exchange}.   
Instead of $J_{j1}$ and $K_{j1}$ with $j>1$, we employ the linear combinations
\begin{equation} \label{tildecoup}
L_{j 1}^{(\pm)} \Big|_{j>1 } = \frac12\left( J_{j 1} \pm K_{j 1} \right).
\end{equation}
The second term in Eq.~\eqref{Hint4} contains contributions that are initially absent, $K_{jj}=0$ for $j>1$, see Eq.~\eqref{exchange}. However, we shall see in Sec.~\ref{sec5b} that such contributions are dynamically  generated during the RG flow.  

We conclude that $H_{\rm eff}=H_{leads}+H_K$ describes a pseudospin coupled to $M$ bosonic modes, where  the pseudospin dynamics encodes quasiparticle poisoning effects in this strongly blockaded Majorana device. Finally, we note that the bosonized description also allows one to incorporate weak electron-electron interactions in the leads in an exact manner \cite{AltlandEgger2013,Beri2013}.  However, we do not consider such effects below.

\subsection{RG equations}\label{sec5b}

We now discuss the RG equations for arbitrary $M$. We have derived them for the bosonized Hamiltonian $H_{\rm eff}=H_{leads}+H_K$, with $H_K$ in Eq.~\eqref{Hint4} and $E_\Delta\to 0$,  by using the operator product expansion technique \cite{Cardy1996}.  We have also confirmed the correctness of the RG equations by an independent derivation using the fermionic representation of $H_{\rm eff}$ in Sec.~\ref{sec3a}. For arbitrary $M$, the closed set of one-loop RG equations, with the indices $j=2,\ldots,M$ and $1<k<j$, is then given by 
\begin{eqnarray}\label{rge1}
\frac{dJ_{j k}}{d\ell} &=& \sum_{n\ne (1,j,k)} J_{jn} J_{nk} + 2\sum_{s=\pm} L_{j 1}^{(s)} L_{k 1}^{(s)} ,\\
\label{rge2}
\frac{ dL_{j 1}^{(\pm)} }{d\ell} &=&\sum_{n\ne (1,j)} J_{j n} L_{n 1}^{(\pm)} \pm 2 
\left( K_{11} - K_{jj} \right) L_{j 1}^{(\pm)} ,\\
\label{rge3}
\frac{dK_{jj}}{d\ell} &=& - \left( L_{j 1}^{(+)} \right)^2 + \left(L_{j1}^{(-)} \right)^2  ,
\\ \label{rge4}
\frac{dK_{11}}{d\ell}&=& - \sum_{n=2}^M \frac{dK_{nn}}{d\ell}. 
\end{eqnarray}
Note that the flow of the couplings $J_{j1}$ and $K_{j1}$ is now contained in the couplings $L_{j1}^{(\pm)}$, see Eq.~\eqref{tildecoup}. The bare (initial) couplings $J_{jk}(\ell=0)$ and $K_{jk}(0)$ have been specified in Eq.~\eqref{exchange}. This equation also determines the $L^{(\pm)}_{j1}(0)$ from Eq.~\eqref{tildecoup}.  Moreover, we always have $J_{kj}=J_{jk}$.  In the above RG equations, we have dropped all RG-irrelevant couplings that are generated during the RG flow but have vanishing initial value.   We mention in passing that for $M=2$, these RG equations become equivalent Eq.~\eqref{M2rge} in Sec.~\ref{sec4}.

Equation \eqref{rge4} implies that $\sum_{n=1}^M K_{nn}(\ell) = K_{11}(0)$, which is dictated by current conservation and stems from the gauge invariance of the system, namely the invariance of the effective action $S_{\rm eff}$ [see Eq.~\eqref{effact} in App.~\ref{appa}] with respect to the simultaneous shift of all $\Phi_j(\tau)$ (where $\tau$ denotes imaginary time) by an arbitrary constant. Noting that the current through the respective tunnel contact is determined by $\langle\partial_\tau \Phi_j\rangle$, current conservation implies $\sum_j \langle\partial_\tau\Phi_j\rangle=0$. Using bosonization identities \cite{Gogolin1998}, this relation is equivalent to $\sum_j \langle\Theta'_j\rangle=0$, and therefore $H_{\rm eff}$ has to be invariant under a uniform shift of all $K_{jj}$. 

Let us now briefly check that known results for the clean TKE are recovered. In the absence of poisoning, which corresponds to removing the Majorana fermion $\eta_1$ by setting $\tilde t=0$, i.e., $K_{jk}=0$ and $L_{j1}^{(\pm)} = J_{j 1}/2$, Eqs.~\eqref{rge1} and \eqref{rge2}  reproduce the RG equations for the TKE \cite{BeriCooper2012},
\begin{equation}\label{TKE}
\frac{dJ_{j k}}{d\ell}  = \sum_{n\ne (j,k)}  J_{j n}J_{n k},
\end{equation}
where $j,k,n=1,\ldots,M$ and $j\ne k$.  For $M>2$, one flows towards an isotropic strong-coupling fixed point \cite{BeriCooper2012,AltlandEgger2013}, 
\begin{equation}\label{TKE2}
J_{jk}\to J (1-\delta_{jk}), \quad \frac{dJ}{d\ell} = (M-2) J^2,
\end{equation}
which represents a non-Fermi liquid quantum critical point of overscreened multi-channel Kondo type  \cite{BeriCooper2012,AltlandEgger2013,Beri2013,Altland2014,Zazunov2014}.  

Interestingly, one can also arrive at the TKE by trading the ``true'' Majorana fermion $\gamma_1$ for the ``poisoning'' Majorana fermion $\eta_1$.  To illustrate this point, let us consider the case $\lambda_1=0$, such that the tunnel coupling between $\gamma_1$ and the attached lead vanishes. For $j>1$,  we then have $J_{j1}=0$ and  $L_{j1}^{(\pm)} = \pm K_{j 1}/2$. Renaming $K_{j1}\to J_{j1}$, the general RG equations [Eqs.~(\ref{rge1})--(\ref{rge4})] again reduce to the TKE equations \eqref{TKE}.  

The presence of a poisoning Majorana fermion ($\eta_1$) in the effective low-energy Hamiltonian $H_{\rm eff}$ has several consequences. First, it implies the opening of $M-1$ additional ``forward scattering'' channels, $\sim\psi^\dagger_j \psi^{}_1 \eta_1 \gamma_j$ in Eq.~\eqref{Hint}, where an electron is transferred from lead $j=1$ to some other lead ($j\ne 1$), and likewise for the conjugate processes.  Second, a new ``backscattering'' channel will open for the poisoned lead, $\sim\psi^\dagger_1 \psi_1 \eta_1 \gamma_1$.  Under the RG flow, this effect also generates additional terms $\sim \psi^\dagger_j \psi^{}_j \tau_z$, i.e., backscattering will appear in the other leads as well.  (Of course, all these processes are subject to the current conservation constraint in Eq.~\eqref{rge4}.)  One can therefore expect a rich interplay between nonlocal features similar to teleportation \cite{Fu2010}, due to forward scattering between different leads, and local backscattering effects within each lead.

As a minimal description capturing the above physics, we now simplify the full set of RG equations [Eqs.~\eqref{rge1}--\eqref{rge4}] by approximating the couplings as follows. 
In our simplified version of the RG equations, we assume that there are only four independent couplings, denoted by $J$, $L_\pm$ and $K$ below. With indices $j=2,\ldots, M$ and $1<k<j$, the exchange couplings are expressed as 
\begin{eqnarray}\label{fourpar}
J_{j k}  &=& J_{kj} = J ,  \quad  L_{j 1}^{(\pm)} = L_\pm / \sqrt{2}, \\ \nonumber 
K_{j j} &=& \tilde{K} ,\quad    K_{11} =\tilde K  + K/2. 
\end{eqnarray}
The initial values for $J, L_+$ and $K$ obtained from Eq.~\eqref{exchange} are positive, with $L_+(0)>|L_-(0)|$.
We here assume the same coupling $K_{jj}=\tilde K$ for each ``unpoisoned'' ($j>1$) lead, with initial value $\tilde K(0)=0$.  Writing $K_{11}=\tilde K+K/2$ as in Eq.~\eqref{fourpar}, we observe that $\tilde K$ does not scale under the RG because of current conservation, see Eq.~\eqref{rge4}, and therefore plays no role in what follows.  
Inserting Eq.~\eqref{fourpar} into Eqs.~\eqref{rge1}--\eqref{rge3}, we arrive at the simplified RG equations, 
\begin{eqnarray} \label{fourrg}
\frac{dJ}{d\ell} & = & (M-3) J^2 + L_+^2 + L_-^2 , \\
\nonumber
\frac{dL_{\pm}}{d\ell} & = & \left[ (M-2) J \pm K \right] L_\pm , \\
\nonumber
\frac{dK}{d\ell} & = & M \left(L_+^2 - L_-^2 \right) .
\end{eqnarray}

As a first check, let us briefly verify that Eq.~\eqref{fourrg} correctly captures the expected TKE in the clean limit, see Eqs.~\eqref{TKE} and \eqref{TKE2}.  In the absence of poisoning, we have $K_{jk}= 0$, resulting in $K= 0$ and $L_+ = L_-= J/\sqrt{2}$.  
We then readily obtain $dJ/d\ell=(M-2)J^2$ from Eq.~\eqref{fourrg}, in accordance with Eq.~\eqref{TKE2}.  
A second check comes from comparing the results of a numerical integration of the full RG equations [Eqs.~\eqref{rge1}-\eqref{rge4}] to the corresponding predictions obtained from the simplified RG equations.  We present this comparison in App.~\ref{appb}, which shows that for $M>3$, the simplified description is justified.  
For $M=3$, the RG flow towards isotropic couplings as expressed by Eq.~\eqref{fourpar} is not yet established.  

We next observe that Eq.~\eqref{fourrg} implies a constant growth of the ratio $L_+/L_-$ during the RG flow, $d\ln(L_+/L_-)/d\ell=2K>0$. Noting also that the couplings $J, K$ and $L_+$ grow, we see that $L_-$ is dynamically suppressed against these couplings.  Neglecting $L_-$ in Eq.~\eqref{fourrg} and considering the case $M\gg 1$, the RG equations~\eqref{fourrg} simplify to
\begin{equation}\label{largeM}
\frac{dJ}{d\ell}=M J^2, \quad \frac{dK}{d\ell}= M L_+^2, \quad \frac{dL_+}{d\ell} = M J L_+.
\end{equation}
These equations predict $dL_+/dJ= L_+/J$, i.e., $L_+\sim J$, which in turn implies that $dJ/dK=(J/L_+)^2\sim 1$. We conclude that all three couplings scale uniformly towards a strong-coupling regime, $K \sim J \sim L_+$, where one eventually leaves the validity regime of the perturbative RG approach. The analysis in App.~\ref{appb} shows that the above suppression of $L_{j1}^{(-)}$ couplings also takes place for $M=3$.  This 
suppression is then followed by a flow towards isotropic couplings (this happens only for $M>3$). 
     
To conclude this section, the perturbative RG equations indicate that the clean TKE found for $M>2$ will be destroyed by ``dangerous'' quasiparticle poisoning. Nevertheless, see App.~\ref{appb}, a simplified four-parameter description is sufficient for $M>3$, where one coupling $(L_-)$ turns out to be irrelevant.  This dynamical suppression of $L^{(-)}_{j1}$ couplings also holds for $M=3$, while then isotropy is not reached.  We exploit these insights now when turning to the strong-coupling regime.

\section{Strong coupling regime} \label{sec6}

Let us now discuss the physics for $M>2$ encountered at very low temperatures, where the strong-coupling regime is approached.   We begin our analysis with the case $M>3$, where it is justified to employ the 
isotropic couplings in Eq.~\eqref{fourpar}, and later return to the case $M=3$. 

For $M>3$, the coupling $L_-$ decouples and can be neglected, while the remaining couplings [$J, K$ and $L_+$ in Eq.~\eqref{fourpar}] become isotropic and simultaneously approach the strong-coupling regime, see Sec.~\ref{sec5b} and App.~\ref{appb}.  Using the shorthand notation \eqref{shorthand}, the exchange interaction in Eq.~\eqref{Hint4} then takes the form \cite{footnote1}
\begin{eqnarray}\label{Hint5}
H_K &=& J \sum_{1<k<j}^M \cos\left( \Phi_j-\Phi_k\right) +\frac{K}{2}\tau_z \Theta_1^\prime \\ \nonumber & + &
\frac{L_+}{\sqrt{2}} \left( \tau_+ \sum_{j=2}^M e^{i(\Phi_j-\Phi_1)} + {\rm h.c.}\right).
\end{eqnarray}
We now perform a unitary transformation, 
\begin{equation}\label{unitary2}
\tilde H_{\rm eff}= e^{-i\frac{K}{2v_F} \tau_z\Phi_1} H_{\rm eff} e^{i\frac{K}{2v_F}\tau_z\Phi_1},
\end{equation}
in order to gauge away the $\Theta_1'$ term. As a result, the exchange term in $\tilde H_{\rm eff}=H_{leads}+\tilde H_K$ takes the form
\begin{eqnarray}\label{Hint6}
\tilde H_K &=& J \sum_{1<k<j}^M \cos\left( \Phi_j-\Phi_k\right)  \\ \nonumber & + &
\frac{L_+}{\sqrt{2}} \left( \tau_+ \sum_{j=2}^M e^{i[\Phi_j-(1+K)\Phi_1]} + {\rm h.c.}\right).
\end{eqnarray}

The strong-coupling regime is now accessible by (i) diagonalizing $\tilde H_K$ for static field configurations $\{\Phi_j\}$, (ii) minimizing the corresponding ground-state energy $E_K[\{\Phi_j\}]$, and (iii) subsequently taking into account quantum fluctuations (caused by $H_{leads}$) around the minimizing field configurations, cf.~Refs.~\cite{AltlandEgger2013,Zazunov2014} for a discussion of this approach in the clean (unpoisoned) case. Step (i) yields the effective exchange energy
\begin{eqnarray}\label{effene}
E_K &=& J \sum_{1<j<k}^M \cos\left( \Phi_j-\Phi_k\right) - \sqrt{\frac{M-1}{2}} L_+\\ \nonumber
&\times& \left [1+ \frac{2}{M-1}\sum_{1<k<j}^M \cos\left( \Phi_j-\Phi_k\right) \right]^{1/2} .
\end{eqnarray}
Note that $E_K$ is invariant under a uniform shift of all fields $\Phi_j$, reflecting charge quantization on the Majorana-Cooper box \cite{AltlandEgger2013,Zazunov2014}. 
Importantly, Eq.~\eqref{effene} is independent of $\Phi_1$, i.e., the poisoned lead decouples from the problem, and transport at low energy scales 
through the corresponding contact $j=1$ will be blocked.  

The low-energy theory for the remaining $M-1$ leads with $j=2,\ldots,M$, which are not attached to the ``poisoned'' tunnel contact, can then be described as TKE with symmetry group SO($M-1$), instead of the unpoisoned case with SO($M$). Indeed, an expansion of the square root  in Eq.~\eqref{effene} is possible for $M\gg 1$ and yields 
\begin{equation}\label{effene2}
E_K\simeq J_{\rm eff} \sum_{1<j<k}^M \cos\left(\Phi_j-\Phi_k\right),
\end{equation}
with $J_{\rm eff}= J-L_+/\sqrt{2(M-1)}$.  Equation \eqref{effene2} is precisely the effective exchange energy describing the ``clean'' TKE for the remaining $M-1$ unpoisoned leads. For $M\gg 1$, the effective exchange coupling $J_{\rm eff}$ is positive and flows towards strong coupling.   We note in passing that the derivation above does not crucially rely on the initial isotropy of the couplings $J$ and $L_+$, but only on the decoupling of $L_-$.  Isotropy then is in fact automatically generated from the effective TKE flow, cf.~Eqs.~\eqref{TKE} and \eqref{TKE2}, emerging in the ``clean'' (unpoisoned) sector.   

We can now directly apply the analysis of Refs.~\cite{AltlandEgger2013,Zazunov2014} for the unpoisoned case after the replacement $M\to M-1$.  Effectively, the poisoning Majorana fermion $\eta_1$ and the ``true'' Majorana fermion $\gamma_1$ do not contribute to the low-energy sector anymore near the strong-coupling fixed point, and the system then represents a ``clean'' TKE with symmetry group SO$(M-1)$.    
For lead indices $2\le j\ne  k\le M$, the conductance $G_{jk}$ between the respective leads follows at $T=0$ as \cite{BeriCooper2012,Zazunov2014}
\begin{equation}\label{gm3}
 G^{(M>3)}_{jk} = \frac{2e^2}{h}  \frac{1}{M-1}.
\end{equation}
Finite-temperature corrections are given by power laws, see Refs.~\cite{BeriCooper2012,Zazunov2014}. Note that transport involving lead 1 is completely blocked due to poisoning, i.e., $G^{(M>3)}_{j1}=0$ for $j>1$.  The above scenario is expected to apply to all cases with $M>3$.

Let us finally return to the case $M=3$, where App.~\ref{appb} shows that the couplings $L_{j1}^{(-)}$ appearing in the Hamiltonian, see Eqs.~\eqref{Hint4} and \eqref{tildecoup}, still become dynamically suppressed during the RG flow and can therefore be dropped in the strong-coupling analysis.  Performing the same unitary transformation as in Eq.~\eqref{unitary2}, we arrive at 
\begin{eqnarray}\label{Hint7}
\tilde H^{(M=3)}_K &=& J\cos\left( \Phi_2-\Phi_3\right)  \\ \nonumber & + &
\left( \tau_+ \sum_{j=2,3}  L_{j1}^{(+)} e^{i[\Phi_j-(1+K)\Phi_1]} + {\rm h.c.}\right).
\end{eqnarray}
Repeating the subsequent steps, we find the effective exchange energy
\begin{eqnarray}\label{effene3}
&& E^{(M=3)}_K =J \cos\left( \Phi_2-\Phi_3\right)   \\ \nonumber
&& - \left [\left(L_{21}^{(+)}\right)^2+\left(L_{31}^{(+)}\right)^2
+2L_{21}^{(+)} L_{31}^{(+)}\cos\left( \Phi_2-\Phi_3\right) \right]^{1/2} .
\end{eqnarray}
We observe that lead 1 again decouples, i.e., transport involving lead 1 is blocked as for the case $M>3$.  In contrast to the latter case, however, no TKE can develop for the two remaining leads (we recall that the TKE requires at least three leads). Since the effective energy in Eq.~\eqref{effene3} pins the phase difference $\Phi_2-\Phi_3$, the conductance $G_{23}$ will now be strongly suppressed. Formally, the situation is identical to a Majorana single-charge transistor under Coulomb
valley conditions \cite{Hutzen2012}. As a consequence,  a tiny residual conductance due to elastic cotunneling may be found at $T=0$, which arises from $2\pi$ slips of the phase difference $\Phi_2-\Phi_3$.
 
\section{Concluding remarks}\label{sec7}

 To conclude, we have discussed a realistic model for quasiparticle poisoning in Coulomb blockaded Majorana devices. For $M=2$ attached leads, the presence of  a ``dangerous'' quasiparticle state (i.e., of very low energy and located close to a tunnel contact) will generate conventional Kondo physics, which in turn could be observed through transport measurements as a Kondo ridge that appears in all Coulomb valleys.  For $M>2$ leads, we have shown that the TKE of the clean system is destabilized by such dangerous quasiparticles.  For $M>3$ and low temperatures, the poisoned system realizes a TKE for the $M-1$ leads not attached to the poisoned tunnel contact, but transport involving the ``poisoned'' lead is blocked.  Furthermore, for $M=3$ leads, the system is predicted to scale towards a decoupled fixed point, where the conductance  between different leads is exponentially small at low energies.  The fundamental difference between the $M=2$ and $M>2$ cases comes from the fact that transport through the box necessarily has to proceed through the poisoned lead ($j=1$) for $M=2$.  For $M>2$, the system instead flows to a fixed point where the poisoned lead
 decouples, and one arrives at the TKE with $M\to M'=M-1$ as long as $M'>2$.
 The case $M=3$ is therefore special, since one ends up with $M'=2$ and the TKE cannot develop anymore. Effectively, one then arrives at a Majorana single-charge transistor, where transport is blocked under valley conditions \cite{Hutzen2012}.

The ``dangerous'' quasiparticle poisoning mechanism discussed in this paper is due to  sub-gap states localized near a tunnel contact. However, the quasiparticle Hamiltonian in Eq.~\eqref{Hqp} also contains delocalized quasiparticle states above the proximity gap in the TS wires, providing yet another source of poisoning.  This effect is important for weakly blockaded systems, where the proximity gap exceeds the charging energy, $\Delta_w>E_C$. In general, a freely propagating quasiparticle is then simultaneously tunnel-coupled to several leads, with amplitudes $\tilde \lambda_j$, generating direct inter-lead tunnel couplings in $H_t$.  Schematically, they have the form 
\begin{equation}\label{Hdirect}
H_{t,{\rm direct}}= \sum_{j\ne k}\frac{\tilde\lambda_{j}\tilde\lambda_k}{\Delta_w} \psi_j^\dagger(0) \psi^{}_k(0) + {\rm h.c.}
\end{equation}
Such terms are expected to be important for pairs of leads ($j\ne k$) that are coupled to the same TS wire. Non-local transport mediated by these states may then compete with tunnel processes via MBSs or subgap quasiparticle states. However, we do not expect dramatic changes to the scenario outlined in this work since Eq.~\eqref{Hdirect} does not generate RG-relevant terms.  

Finally, let us note that we did not include a direct tunnel coupling, $t_d$,
between the ``true'' Majorana fermion $\gamma_1$ and the poisoning Majorana
fermion $\eta_1$, which would give a contribution $H_d = i t_d \gamma_1 \eta_1$.
Such a term could arise from the direct hybridization of $\gamma_1$ and the
quasiparticle state, and then leads to an effective Zeeman field for the
enlarged ``spin'' formed by $\eta_1$ and the $\gamma_j$.
When the quasiparticle corresponds to an eigenstate of the TS wire, however,
it is by definition orthogonal to the MBS, and $t_d=0$.  Only under
``extrinsic'' quasiparticle poisoning,  $t_d\ne 0$ is possible, which will then act as magnetic field for the spin-$1/2$ Kondo effect for $M=2$. For $M>2$ and low
temperatures, however, since both $\eta_1$ and $\gamma_1$ decouple from the low-energy sector, such a coupling does not affect the physics.

We hope that our study will be helpful to the interpretation of experiments on Coulomb-blockaded Majorana devices as well as to future theoretical studies of related questions. 
  
\acknowledgments{We thank A. Altland, B. B{\'e}ri, K. Flensberg, C.M. Marcus, E. Sela, and A. Levy Yeyati for useful discussions. This work has been supported by the Deutsche Forschungsgemeinschaft within network SPP 1666 (R.E.) and by a Humboldt Prize of the Alexander-von-Humboldt foundation, enabling an extended stay of A.M.T. in D\"usseldorf.}

\appendix
\section{Bosonization and parity operators}\label{appa}

We here discuss the bosonized representation of the exchange interaction $H_K$ in Eq.~\eqref{Hint}, see Sec.~\ref{sec5a}. Let us start by defining a set of parity operators (the superscript refers to the respective tunnel contact),
\begin{equation}
\sigma_z^{(j)} = 2 i \Gamma_j \gamma_j ,\quad \sigma_x^{(1)} = 2 i \Gamma_1 \eta_1 ,
\quad \sigma_y^{(1)} = 2 i \eta_1 \gamma_1 ,
\end{equation}
which combine physical ($\gamma_j$, $\eta_1$) and auxiliary ($\Gamma_j$) Majorana fermions, cf.~Ref.~\cite{Altland2014}.  Several comments are in order at this point: 
\begin{enumerate}[(i)]
\item Each of the above parity operators is Hermitian and squares to unity, i.e., it has eigenvalues $\pm 1$. 
\item All $\sigma_z^{(j)}$ operators are mutually commuting, and the $\sigma_z^{(j\ne 1)}$ operators commute with $\sigma_{x,y}^{(1)}$. 
\item The three operators $\sigma_{a=x,y,z}^{(j=1)}$, which refer to the tunnel contact where the poisoning quasiparticle is located,
obey the commutator rules of Pauli matrices, $\left[\sigma_a, \sigma_b\right] = i \epsilon_{abc}\sigma_c,$ where $\epsilon_{abc}$ is the antisymmetric 3D Levi-Civit{\`a} tensor and we sum over double indices.  
 \item The charge quantization condition $Q=[n_g]$ resulting from Eq.~(\ref{cond1}) is equivalent to a total fermion parity constraint on the box \cite{BeriCooper2012,Altland2014}. This implies that the fermion parity operator, $P\sim \eta_1\prod_j\gamma_j$, must be conserved, $\left [H_{\rm eff}, P\right]=0$. 
\item Products of $\sigma_z$ operators for $j\ne 1$ and $k\ne 1$ define conserved quantities $p_{jk}=\sigma_z^{(j)} \sigma_z^{(k)} =\pm 1$, which commute with both $H$ and $P$.  Since $p_{jk}=\pm 1$ implies equivalent physical results after suitable $\pi$-shifts of 
boson fields \cite{Altland2014,Zazunov2014}, we are free to put, say, $p_{jk}=+1$.   
\end{enumerate}
Using the shorthand notation \eqref{shorthand}, we obtain the bosonized form of the exchange interaction (\ref{Hint}) as 
\begin{eqnarray}\label{HintA}
H_{K}&=& \sum_{1<k<j}^M J_{jk} \cos \left( \Phi_j - \Phi_k \right)
\\ \nonumber &+&\sum_{j = 2}^M \sigma_z^{(j)} \Biggl[ \sigma_z^{(1)} J_{j1} \cos \left( \Phi_j - \Phi_1 \right) \\
\nonumber &+& \sigma_x^{(1)} K_{j1} \sin \left( \Phi_j - \Phi_1 \right) \Biggr] - K_{11} \sigma_y^{(1)} \Theta'_1 .
\end{eqnarray}
At this stage, we note that for $j\ne 1$ and $k\ne 1$, the commutation relation $\left[ \sigma_z^{(j)} \sigma_a^{(1)}, \sigma_z^{(k)} \sigma_b^{(1)} \right] =i \epsilon_{abc} \sigma_c^{(1)}$ holds. As a consequence, we may equivalently replace the parity operator products appearing in Eq.~\eqref{HintA} by a new set of Pauli matrices $\tau_{a}$.  For $j\ne 1$ and $a=x,z$, we thus put 
$\sigma_z^{(j)} \sigma_a^{(1)} \to \tau_a$, and  rename $\sigma_y^{(1)}\to \tau_y$. 
This step does not involve any approximation but serves to simplify the problem. We thereby arrive at
\begin{eqnarray}\label{Hint3}
 H_{K} &= &\sum_{1<k<j}^M J_{jk} \cos \left( \Phi_j - \Phi_k \right) - K_{11} \tau_y \Theta'_1\\
 \nonumber &+&\sum_{j = 2}^M \left[ J_{j 1} \tau_z \cos \left( \Phi_j - \Phi_1 \right) + K_{j1} \tau_x\sin \left( \Phi_j - \Phi_1 \right)  \right].
 \end{eqnarray}
 Finally, for the derivation of the RG equations in Sec.~\ref{sec5b}, we found it convenient to perform a final unitary transformation of $H_{\rm eff}$. After two consecutive $\pi/2$-rotations in $\tau$-space, $\left( \tau_y, \tau_z \right) \to \left( -\tau_z, \tau_y \right)$ and $\left( \tau_x, \tau_y \right) \to \left( -\tau_y, \tau_x \right),$ and taking into account that $J_{jk}=J_{kj}$ and $K_{jk}$ are real-valued, see Eq.~\eqref{exchange}, we obtain Eq.~\eqref{Hint4} quoted in the main text.

By integration over the Gaussian boson fields at $x<0$, the Euclidean action corresponding to $H_{\rm eff}$ can be expressed in term of boundary boson fields, $\Phi_j(\tau)$ and $\Theta'_1(\tau)$, with $\tau$ denoting imaginary time, cf.~Eq.~\eqref{shorthand}, plus an additional triplet of Majorana fermion fields, $\xi_{a=x,y,z}(\tau)$. This Majorana triplet captures the pseudospin dynamics according to the standard relation $\tau_a = 2 i \epsilon_{a b c} \xi_b \xi_c.$
After some algebra, with $\Phi_j(\tau)=T\sum_\omega e^{i\omega \tau} \tilde\Phi_j(\omega)$ expressed as a Fourier series involving the bosonic Matsubara frequencies  $\omega$, we find the Euclidean action 
\begin{eqnarray}\label{effact}
S_{\rm eff} &=& \sum_{\omega}  \frac{T|\omega|}{2\pi} \sum_{j=1}^M \left| \tilde\Phi_j(\omega) \right|^2\\  \nonumber
&+& \int_0^{1/T} d \tau \left( \frac12 \sum_{a} \xi_a \partial_\tau \xi_a + H_{K}(\tau) \right) ,
\end{eqnarray}
where the first term describes Ohmic dissipation due to electron-hole pair excitations in the leads \cite{Gogolin1998}.  

\section{On the RG equations }\label{appb}

In this appendix, we shall discuss several details concerning Sec.~\ref{sec5b} on the RG equations for $M>2$.  

First,  we will show that always one of the sets of couplings $L^{(+)}_{j1}$ and $L^{(-)}_{j1}$, see Eq.~\eqref{tildecoup}, becomes subleading after an initial transient of the RG flow.  
In a slight abuse of terminology, we call this behavior ``helicity,'' since thereby a specific  coupling mechanism between the ``spin'' operators $\tau_\pm$ and the leads will be selected in Eq.~\eqref{Hint4}. When following the RG flow towards the strong-coupling regime, one can then neglect these subleading couplings.  We note that in Secs.~\ref{sec5b} and \ref{sec6}, it has been assumed that the $L^{(-)}_{j1}$ couplings are subleading, which is in accordance with the initial conditions for the exchange couplings in Eq.~\eqref{exchange}. However, one can check that identical results for observables also follow in the opposite case.  

In a second step, we then demonstrate that the simplified RG equations \eqref{fourrg} provide a useful approximation for the full RG equations when $M>3$.  We call this phenomenon ``isotropization,'' since the full set of couplings can then effectively be replaced by just four isotropic couplings in Eq.~\eqref{fourpar}. Since helicity always sets in before isotropization (see below), the coupling $L_-$ can be neglected in addition, and one has only three relevant RG couplings for all $M>3$.  For $M=3$, on the other hand, we find that the system becomes helical but not isotropic during the RG flow.

\begin{figure}[t]
  \centering  
  \includegraphics[width=8.2cm]{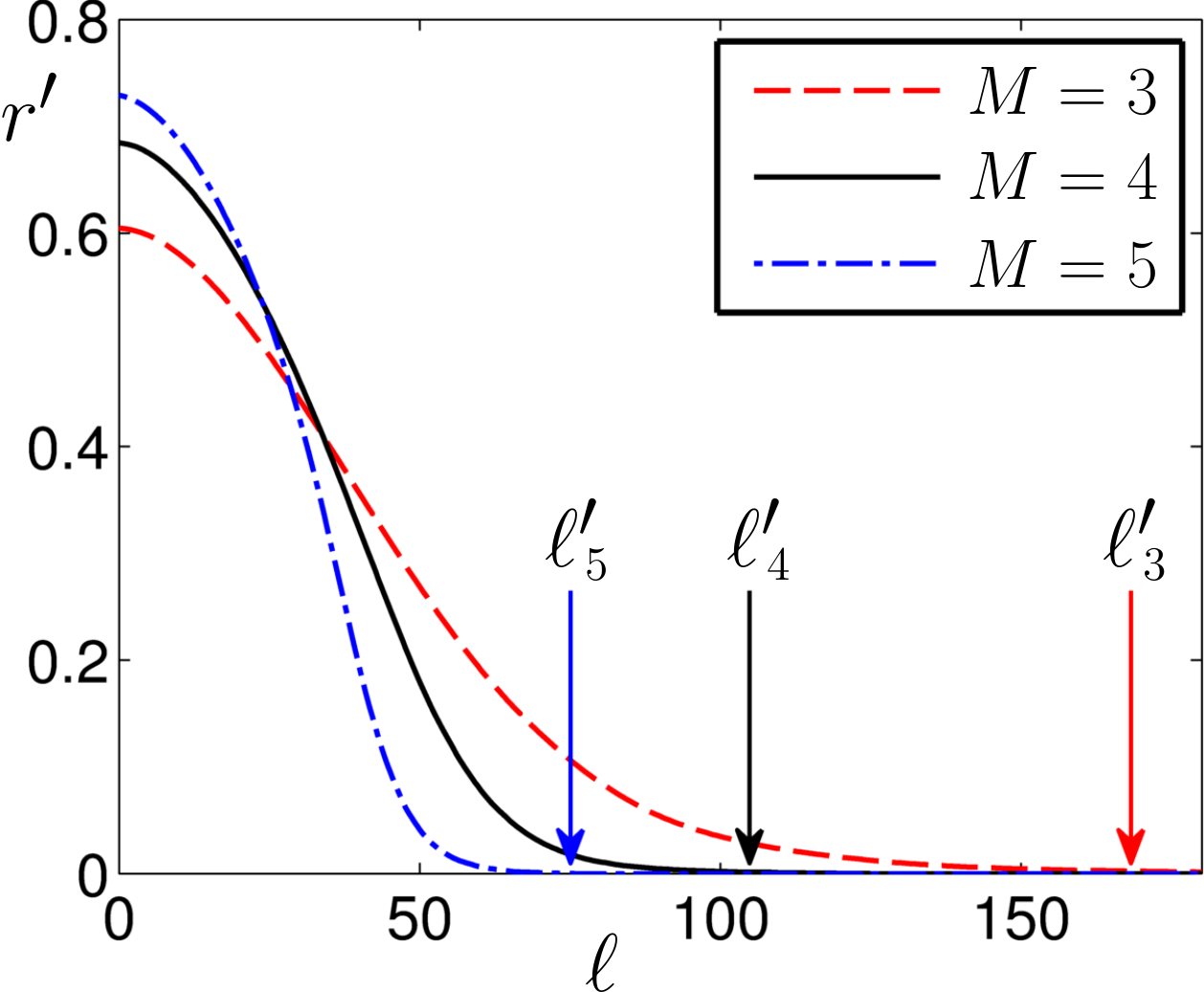}
  \caption{(Color online) RG flow of the helicity parameter $r'$ vs $\ell$, see Eq.~\eqref{decoupling}, for $M=3, 4$ and $5$.  Each curve has been obtained by numerical solution of the full RG equations [Eqs.~\eqref{rge1}--\eqref{rge4}], taking an average over $10^{4}$ random initial configurations   \cite{foot3}.  The arrows labelled with 
$\ell'_{M}$ indicate the flow parameter values at which $r'(\ell')=0.01$ is reached for the respective ($M=3,4,5$) curve. 
  } \label{fig2}
\end{figure}

To study these questions, we have numerically solved the full RG equations [Eqs.~\eqref{rge1}-\eqref{rge4}], starting from randomly chosen (but positive) initial values of the exchange coupling matrix elements $J_{jk}$, $L_{j1}^\pm$, and $K_{jj}$. In order to obtain generic initial values, we draw these from a uniform probability distribution \cite{foot3}. From the RG solution at given flow parameter $\ell$, we compute the helicity parameter
\begin{equation}\label{decoupling}
r'(\ell) = 1- \left|\frac{\sum_{j=2}^M\sum_{s=\pm}sL_{j1}^{(s)}} {\sum_{j,s}L_{j1}^{(s)}}\right|,
\end{equation}
where $r'(\ell)\to 0$ indicates fully established helical behavior.   In Fig.~\ref{fig2}, for several values of $M$, we show our results for $r'(\ell)$ obtained by averaging over a large set of initial values \cite{foot3}.  We find helicity for all $M\geq 3$, where the transition to helical behavior becomes faster with increasing $M$.  To quantify this crossover, we employ the scale $\ell'$ for $M=3,4,5$ in Fig.~\ref{fig2}, which is defined such that $r'(\ell')=0.01$.  At lower energy scales ($\ell>\ell'$), helicity is thus well established, with $r'<0.01$.  

\begin{figure}[t]
  \centering  
  \includegraphics[width=8.5cm]{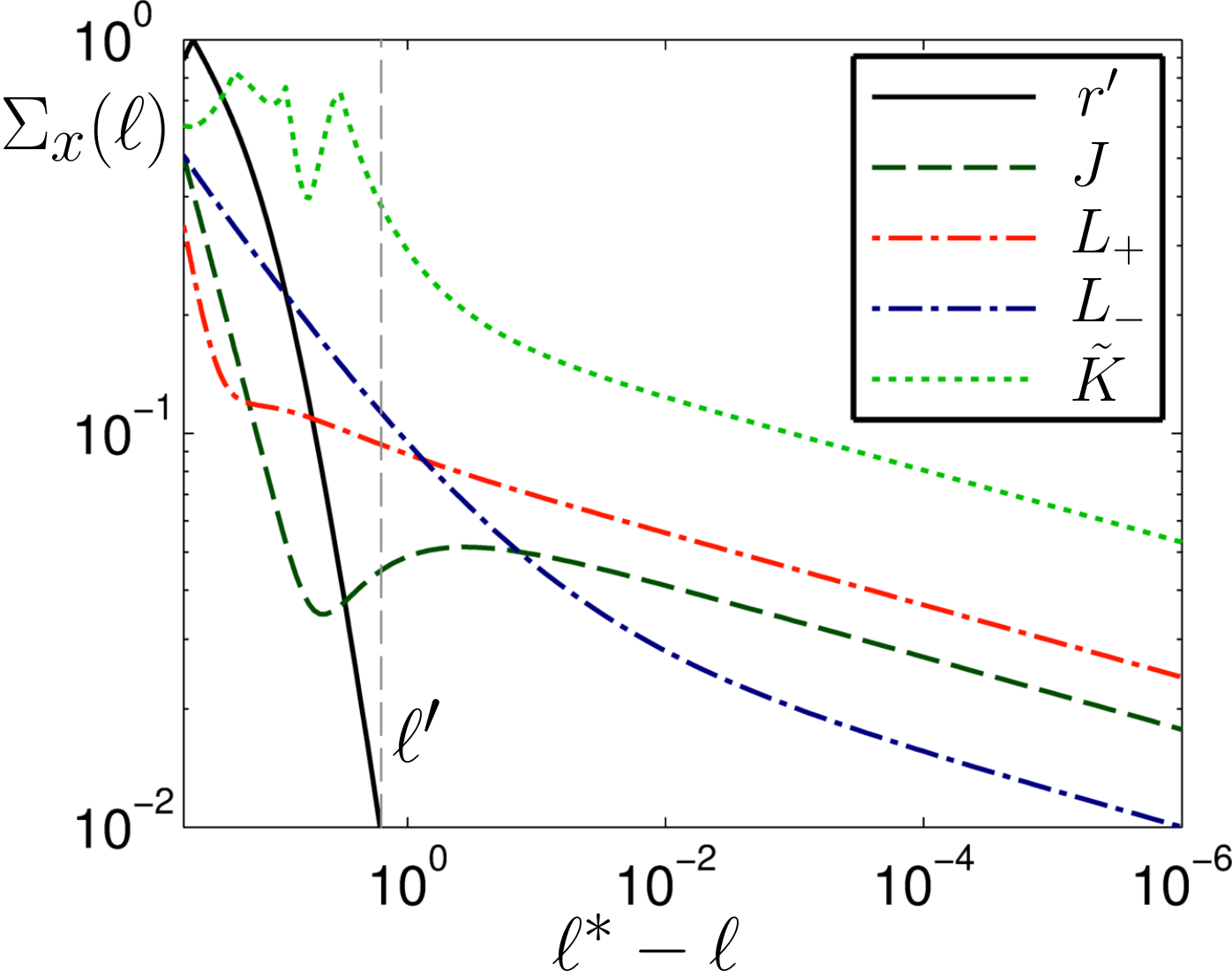}
  \caption{(Color online) RG flow at $M=4$  for the normalized standard deviations  $\Sigma_x(\ell)$ with $x\in \lbrace J, L_+, L_-, \tilde K\rbrace$, see Eq.~\eqref{b2},  and for the helicity parameter $r'$ in Eq.~\eqref{decoupling}, cf.~Fig.~\ref{fig2}. We show all quantities vs $\ell^*-\ell$, where $\ell^\ast$ indicates the point at which the RG solution diverges. For the shown parameters, $\ell^\ast \approx 54.5$. The scale $\ell'$ is defined by $r'(\ell')=0.01$ and approximately indicates the crossover to helical behavior. Here $\ell'\approx \ell^\ast-1$  corresponds to the vertical thin-dashed line. For $\ell>\ell'$, the system subsequently flows to isotropy, $\Sigma_x(\ell>\ell')\to 0$.
  } \label{fig3}
\end{figure}

Next, in order to test for the isotropization of the couplings, we compute the average values 
\begin{eqnarray}\label{avgvalue}
J &=& \frac{2}{(M-1)(M-2)}\sum_{1<j<k}^M J_{jk},\\ \nonumber
 L_\pm &=& \frac{1}{M-1}\sum_{j=2}^M L_{j1}^{(\pm)},\quad \tilde K =\frac{1}{M-1}\sum_{j=2}^M K_{jj}, 
\end{eqnarray} 
and then monitor the RG flow of anisotropy measures, $\Sigma_x(\ell)$, defined for each parameter family $x\in \lbrace J, L_+, L_-,  \tilde{K}\rbrace$. 
The quantities $\Sigma_x$ indicate how well the simplified couplings in Eq.~\eqref{fourpar} can approximate the full RG flow of matrix elements of type $x$. They are defined as standard deviations of $x$-type matrix elements normalized by their respective average value,
\begin{eqnarray}\nonumber
\Sigma_J^2 &=& \frac{2}{(M-1)(M-2)} \sum_{1<j<k}^M \frac{(J_{jk}-J)^2}{J^2} ,\\ \label{b2}
\Sigma_{L_\pm}^2 &=& \frac{1}{M-1}\sum_{j=2}^M  \frac{(L_{j1}^{(\pm)}-L_\pm)^2}{L_\pm^2 }, \\ \nonumber
\Sigma_{\tilde K}^2 &=&  \frac{1}{M-1}\sum_{j=2}^M  \frac{(K_{j1}-\tilde K)^2}{\tilde K^2 }, 
\end{eqnarray}
and thus quantify the geometrical distance of the respective matrix elements from the isotropic case. We take $\Sigma_x(\ell)$ as bona-fide measure for the (an)isotropy of type-$x$ couplings.
In Fig.~\ref{fig3}, we show typical numerical results obtained for $M=4$ \cite{foot2}. We observe that after the system becomes helical, i.e., for $\ell>\ell'$, all anisotropies $\Sigma_x(\ell)$ become more and more suppressed during the RG flow.  This suppression implies isotropic behavior, thereby justifying the simplified RG equations \eqref{fourrg}. The initial fluctuations visible in $\Sigma_{\tilde{K}}(\ell)$ can be rationalized by noting that the $\tilde K$-average in Eqs.~\eqref{avgvalue} and \eqref{b2} does not directly include $K_{11}$, thus allowing for exchange towards $K_{11}$ not captured by $\Sigma_{\tilde{K}}$.  

In our numerical study, we always find that helicity sets in before isotropization.   
Similar calculations as depicted in Fig.~\ref{fig3} have also been performed for $M=5$ (not shown), where the approach to helicity, and subsequently to isotropy, is faster, cf.~also Fig.~\ref{fig2}. This numerical result is consistent with our strong-coupling analysis for $M\gg 1$ in Sec.~\ref{sec6}, where we find a TKE  of SO($M-1$) symmetry for the $M-1$ ``unpoisoned'' leads. Isotropy then follows from Eq.~\eqref{TKE2}, corresponding to a decay of all $\Sigma_x(\ell)$.  In effect, the TKE emerging in the unpoisoned sector thus drives all couplings towards isotropy.  For a ``large'' unpoisoned sector ($M\gg 1$), this isotropization is very rapid.
For $M=3$, on the other hand, the unpoisoned sector does not allow for a TKE and therefore isotropization is absent.


\begin{thebibliography}{99}
\bibitem{footnote1}
Putting $K_{jj} = \tilde K$ for $j > 1$ as in Eq.~\eqref{fourpar}, and exploiting current conservation, Eq.~\eqref{rge4}, we may shift $K_{jj}\to K_{jj}-\tilde K$ for all $j = 1,\ldots, M$ in Eq.~\eqref{Hint4}. Consequently, only the $\Theta_1^\prime$ term, with coupling $K = 2(K_{11} - \tilde K)$, remains in Eq.~\eqref{Hint5}.
\bibitem{foot3}
Initial values for the exchange couplings (in units with $v_F=1$) are chosen as positive uniform random numbers $\in [0.0001,0.0101]$, where in each round we first choose values for $M=5$, and subsequently discard all couplings not needed for $M=4$ and  $M=3$.  The rather large values observed in Fig.~\ref{fig2} for the scale $\ell'$ (at which helicity sets in) are rationalized by noting that our average over initial conditions also includes some ``pathological'' ones with a very slow RG flow towards the strong-coupling regime.
\bibitem{foot2}
In units with $v_F=1$, the initial parameters used in Fig.~\ref{fig3} are:     
$J_{23} = 0.0062, J_{34} = 0.0091, J_{24} =0.0020, L^{(+)}_{21} = 0.0076, L^{(+)}_{31} = 0.0036, L^{(+)}_{41} = 0.0043,
L^{(-)}_{21} = 0.0017, L^{(-)}_{31} = 0.0083, L^{(-)}_{41} = 0.0063, K_{11} = 0.0096, K_{22} = 0.0075, K_{33} = 0.0082$, and $K_{44} = 0.0008$.
However, the precise choice of initial parameters does not affect our general conclusions.
In particular, we could use random tunnel couplings to parametrize the exchange coupling matrices, see Eq.~\eqref{exchange}, with $L_{j1}^{(+)}>L_{j1}^{(-)}$ and $K_{jj}=0$ (for $j>1$). Aside from transients in the RG flow, however, the approach to helicity and subsequently to isotropy does not depend on such details, and one obtains qualitatively similar plots.
\end{thebibliography}
\end{document}